\documentclass[aps,prl,twocolumn,showpacs,groupedaddress]{revtex4}  
\usepackage{graphicx}  
\usepackage{dcolumn}   
\usepackage{bm}        
\usepackage[latin1]{inputenc}
\usepackage{verbatim}
\usepackage[]{epsfig}
\usepackage{psfrag}
\usepackage{fancyhdr}
\usepackage{float}
\usepackage{graphicx}
\usepackage{subfigure}
\usepackage{wrapfig}
\usepackage{sidecap}

\newcommand{\invisible}[1]{}

\RequirePackage{xspace}
\def\b     {\ensuremath{b}\xspace}
\def\bbar  {\ensuremath{\overline b}\xspace}
\def\BR         {{\ensuremath{\cal B}\xspace}}
\newcommand{\gevcc}{\ensuremath{{\mathrm{\,Ge\kern -0.1em V\!/}c^2}}\xspace}
\newcommand{\gevc}{\ensuremath{{\mathrm{\,Ge\kern -0.1em V\!/}c}}\xspace}
\def\bc   {\begin{center}}
\def\ec   {\end{center}}
\newcommand{\jprlBase}       {Phys.\ Rev.\ Lett.\xspace}
\newcommand{\jprl}      [1]  {\jprlBase\ {\bf #1}}

\newcommand{\insertMonoFig}[3] {
\begin{figure}[h]
   \begin{center}
   \includegraphics[width=5.5cm]{#1}
   \caption{#2}
   \label{#3}
  \end{center}
\end{figure}
}

\begin{document}

\hspace{5.2in} \mbox{Fermilab-Pub-08/142-E}

\title{Search for neutral Higgs bosons in multi-$b$-jet events in $p\bar{p}$ collisions at $\sqrt{s} = 1.96$ TeV}
%
\author{V.M.~Abazov$^{36}$}
\author{B.~Abbott$^{75}$}
\author{M.~Abolins$^{65}$}
\author{B.S.~Acharya$^{29}$}
\author{M.~Adams$^{51}$}
\author{T.~Adams$^{49}$}
\author{E.~Aguilo$^{6}$}
\author{S.H.~Ahn$^{31}$}
\author{M.~Ahsan$^{59}$}
\author{G.D.~Alexeev$^{36}$}
% ...

\affiliation{\vspace{0.1 in}(The D\O\ Collaboration)\vspace{0.1 in}}
% ...
\affiliation{$^{6}$University of Alberta, Edmonton, Alberta, Canada,
                Simon Fraser University, Burnaby, British Columbia, Canada,
                York University, Toronto, Ontario, Canada, and
                McGill University, Montreal, Quebec, Canada}
% ...
\affiliation{$^{29}$Tata Institute of Fundamental Research, Mumbai, India}
% ...
\affiliation{$^{31}$Korea Detector Laboratory, Korea University, Seoul, Korea}
% ...
\affiliation{$^{36}$Joint Institute for Nuclear Research, Dubna, Russia}
% ...
\affiliation{$^{49}$Florida State University, Tallahassee, Florida 32306, USA}
% ...
\affiliation{$^{51}$University of Illinois at Chicago,
                Chicago, Illinois 60607, USA}
% ...
\affiliation{$^{59}$Kansas State University, Manhattan, Kansas 66506, USA}
% ...
\affiliation{$^{65}$Michigan State University,
                East Lansing, Michigan 48824, USA}
% ...
\affiliation{$^{75}$University of Oklahoma, Norman, Oklahoma 73019, USA}
% ...
\date{May 22, 2008}

\begin{abstract}
Data recorded by the D0 experiment at the Fermilab Tevatron Collider are analyzed to search for neutral Higgs bosons produced in association with $b$ quarks. This production mode can be enhanced in the minimal supersymmetric standard model (MSSM). The search is performed in the three $b$ quark channel using multijet triggered events corresponding to an integrated luminosity of $1~\mathrm{ fb}^{-1}$. No statistically significant excess of events with respect to the predicted background is observed and limits are set in the MSSM parameter space.
\end{abstract}

\maketitle

Supersymmetry (SUSY)~\cite{mssm} is a popular extension of the standard model (SM) which overcomes the hierarchy problem associated with electroweak symmetry breaking and the Higgs mechanism. In contrast to the SM, where only one Higgs doublet is required to break the SU(2) symmetry, SUSY requires the presence of at least two Higgs doublets. In the MSSM five Higgs bosons remain after electroweak symmetry breaking; three neutral: $h$, $H$, and $A$ - denoted as $\phi$, and two charged: $H^\pm$. The Higgs sector can be parameterized by $\tan\beta$, the ratio of the two Higgs doublet vacuum expectation values, and $m_A$, the mass of the pseudo-scalar Higgs boson $A$.

The Higgs-quark couplings in the MSSM are proportional to their SM counterparts, with the exact factor depending on the type of quark (up- or down-type) and on the type of Higgs boson. For large values of $\tan\beta$ at least two Higgs bosons (either $A$ and $h$, or $A$ and $H$) have approximately the same mass and couplings to down-type quarks, which are enhanced by a factor $\tan\beta$ relative to the SM ones, while the couplings to up-type quarks are suppressed. In this large $\tan\beta$ region the three Higgs boson couplings follow the sum rule $g_{\rm hbb}^2+g_{\rm Hbb}^2+g_{\rm Abb}^2\approx 2\times\tan^2\beta \times g_{\rm h_{SM}}^2$. In $p\bar{p}$ collisions at $\sqrt{s} = 1.96$~TeV at the Fermilab Tevatron Collider, the production of Higgs bosons associated with bottom quarks (highest mass down-type quark) is therefore, in these cases, enhanced by a factor $2\times\tan^2\beta$ relative to the SM. Due to the $\tan\beta$ enhancement, the main decay for all these bosons is $\phi\to \b\bbar$ (the branching fraction, $\BR(\phi\to \b\bbar)$, is $\approx 90\%$).  The enhanced production and branching ratio make the final state with three $b$ jets an important channel in the search for MSSM Higgs bosons at large $\tan\beta$. At a hadron collider this final state has a large background from multijet production which is poorly modeled by simulation, making the search for this topology very challenging.

MSSM Higgs boson production has been studied at LEP which excluded $m_{h,A}<93~\gevcc$ for all $\tan\beta$ values~\cite{cite:LEP_exclu}. CDF~\cite{cite:CDF_exclu,cite:CDF_exclu2} and D0~\cite{cite:D0_exclu, cite:D0_exclu2} have extended the MSSM Higgs boson searches to higher masses for high $\tan\beta$ values. The result presented in this Letter supersedes our previous published result~\cite{cite:D0_exclu}. In addition to including more data, this analysis benefits from improved signal and background modeling and an improved limit setting procedure, which uses only the shape, and not the normalization, of the final discriminating variable. 

The D0 detector is described in Ref.~\cite{run2det}. Dedicated triggers designed to select events with at least three jets are used in this analysis. Typical requirements are at least two jets with transverse momenta $p_T > 25$ \gevc, an additional jet with  $p_T > 15$ \gevc, and the $p\bar{p}$ interaction vertex is required to be reconstructed well within the geometric acceptance of the silicon detector. Algorithms for identifying $b$ jets at the trigger level are also employed in about 70\% of the integrated luminosity used for this analysis. After data quality requirements the total data sample corresponds to $1.02 \pm 0.06~\mathrm{ fb}^{-1}$~\cite{d0lumi}.

Signal samples are generated for Higgs boson masses from $90$-$220$~\gevcc using the leading order {\sc pythia} event generator~\cite{pythia} to generate associated production of $\phi$ and a $b$ quark in the 5-flavor scheme, $gb\rightarrow \phi b$. Weights, calculated with {\sc mcfm}~\cite{cite:MCFM}, are applied to the signal samples as a function of $p_T$ and $\eta$ of the leading $b$ jet which is not from the decay of the Higgs boson, to correct the cross section and experimental acceptance to next-to-leading order (NLO). Multijet background events from the $b\bar{b}$, $b\bar{b}j$, $b\bar{b}jj$, $c\bar{c}$, $c\bar{c}j$, $c\bar{c}jj$, $b\bar{b}c\bar{c}$, and $b\bar{b}b\bar{b}$ processes (where $j$ denotes a light parton: $u$, $d$, $s$ quark or $\rm gluon$) are generated with the {\sc alpgen}~\cite{alpgen} event generator. The contributions from other processes, such as $t\bar{t}$, $Zb\bar{b}$, and single top production, are found to be negligible. The {\sc alpgen} samples are processed through {\sc pythia} for showering and hadronization. All samples are then processed through a {\sc geant}-based \cite{geant} simulation of the D0 detector and the same reconstruction algorithms as the data. A parameterized trigger simulation is used to model the effects of the trigger requirements on the simulated events. 

Jets are reconstructed from energy deposits in calorimeter towers using the midpoint cone algorithm~\cite{cone} with radius $= 0.5$. Jet reconstruction and energy scale determination are described in detail in Ref.~\cite{cite:jetx}. All calorimeter jets are required to pass a set of quality criteria with about 98\% efficiency and have at least two reconstructed tracks within $\Delta {\cal R}($track, jet-axis$)=\sqrt{(\Delta\eta)^2 +(\Delta\varphi)^2}<0.5$ (where $\eta$ is the pseudorapidity and $\varphi$ the azimuthal angle). 

We select signal events by requiring at least three and at most five jets with $p_T > 20$ \gevc and $|\eta| < 2.5$. A neural network (NN) based $b$-tagging algorithm~\cite{nnbtag2}, with lifetime based information involving the track impact parameters and secondary vertices as inputs, is used to identify $b$ jets. Each event must have at least three jets satisfying a tight $b$-tag NN requirement. The single jet $b$-tagging efficiency is $\approx$~50$\%$ for a light-jet mistag rate of $\approx$~0.4$\%$. The events with at least two tight $b$-tags are also kept and used to model the background. Simulated events are weighted based on their tagging and fake rate probabilities determined from data. Finally, the transverse momenta of the two highest $p_T$ jets which are also $b$-tagged are required to be above $25$ \gevc. To further increase the sensitivity, the analysis is split into separate three-, four-, and five-jet channels. After the event selection 3,224 events remain in the exclusive three-jet sample, 2,503 and 704 events, respectively, in the four- and five-jet sample. The signal efficiencies for Higgs boson masses between $100$ and $200$ \gevcc range from $0.3-1.2\%$ in the three-jet channel ($0.2-0.6\%$ and $0.01-0.12\%$ in the four- and five-jet channels). 

The background composition is determined separately for each jet multiplicity. The fractional contribution $\alpha_{i}$ of the $i$th background process is calculated from equations linking the $b$-tag efficiency, $\epsilon_{j}$, in an event with the $N_j$ observed events:
\begin{equation}
 \begin{array}{r c l}
  \sum_{\displaystyle i} \alpha_{i}                      &=& 1 \vspace{2mm}\\
  \sum_{\displaystyle i} \alpha_{i} \times\ \epsilon_{j}^{i}& =& N_{j}/N_{\rm tot}
  \end{array}
\label{eq:system}
\end{equation}
Here, $j$ indicates single, double and triple $b$-tagged jets in an event for different $b$-tag criteria, and $N_{\rm tot}$ is the total number of events. The double $b$-tagged sample is found to be dominated by $b\bar{b}j$ while the triple $b$-tagged sample consists of a mix of $\approx 50\%$ $b\bar{b}b$, $\approx 30\%$ $b\bar{b}j$, and $\approx 20 \%$ $b\bar{b}c+bc\bar{c}$. An alternative method to determine the background, based on fitting simulated $H_T = \sum p_{T\mathrm{jet}}$ shape templates to the data, confirms the composition of the background.

For every event the two jet pairs with the largest summed transverse momenta are considered as possible Higgs boson candidates. To remove discrepancies between data and simulation originating from gluon splitting ($g\rightarrow b\bar{b}$), only jet pairs with $\Delta {\cal R} > 1.0$ are considered in the final analysis. 

The following six variables separate the signal from the backgrounds and are well modeled by the simulation: the difference in pseudorapidity between the two jets in the pair; the azimuthal angular difference between the two jets in the pair; the angle between the leading jet in the pair and the total momentum of the pair; $|\vec{p_{b_1}} - \vec{p_{b_2}}|/|\vec{p_{b_1}} +\vec{p_{b_2}}|$, the momentum balance in the pair; the rapidity of the pair; and the event sphericity. Based on these kinematic variables, a likelihood discriminant $\mathcal{D}$, is calculated according to:
\begin{equation}
    \mathcal{D}(x_1,....,x_6) =\frac{\prod_{i=1}^{6}{ p^{sig}_i (x_i)}}{\prod_{i=1}^{6}{ p^{sig}_i (x_i)}+\prod_{i=1}^{6}{ p^{bkg}_i (x_i)}},
\end{equation}
where $p^{sig}_i$ ($p^{bkg}_i$) refers to the signal (background) probability density function (pdf) for variable $x_i$, and $(x_1,...,x_6)$ is the set of measured kinematic variables for the jet pair. The pdfs are obtained from triple $b$-tagged signal and background simulation. Two likelihoods are built combining simulated samples in the $90-130$\gevcc (``Low-mass'') and $130-220$\gevcc (``High-mass'') mass ranges, providing discrimination at low and high masses, respectively. Studies show that this division of the mass range gives the best discrimination.

Several multijet processes contribute to the background and the uncertainty on the cross sections is large. The $bbb$ component may also contain a contribution that is indistinguishable from a signal and cannot be normalized from the data. To model the background we therefore rely on a combination of data and simulation. The distribution of the expected triple $b$-tagged (3Tag) sample in the two-dimensional $\mathcal{D}$ and invariant mass ($M_{bb}$) plane, $S_{\rm 3Tag}^{exp}(\mathcal{D},M_{bb})$, is obtained from the double $b$-tagged (2Tag) data shape multiplied by the ratio of the simulated (MC) shapes of the triple and double tagged events:
\begin{equation}
S_{\rm 3Tag}^{exp}(\mathcal{D},M_{bb}) =\frac{S_{\rm 3Tag}^{MC}(\mathcal{D},M_{bb})}{S_{\rm 2Tag}^{MC}(\mathcal{D},M_{bb})}S_{\rm 2Tag}^{data}(\mathcal{D},M_{bb}).
   \label{eq:BkgModel}
\end{equation}
Many uncertainties affecting the simulation cancel in the ratio $\frac{S_{\rm 3Tag}^{MC}(\mathcal{D},M_{bb})}{S_{\rm 2Tag}^{MC}(\mathcal{D},M_{bb})}$. Figure~\ref{fig:ThreeTagModel_YLH_MH100} shows $\mathcal{D}$ for data and background for the low-mass likelihood in the three-jet channel.
\insertMonoFig{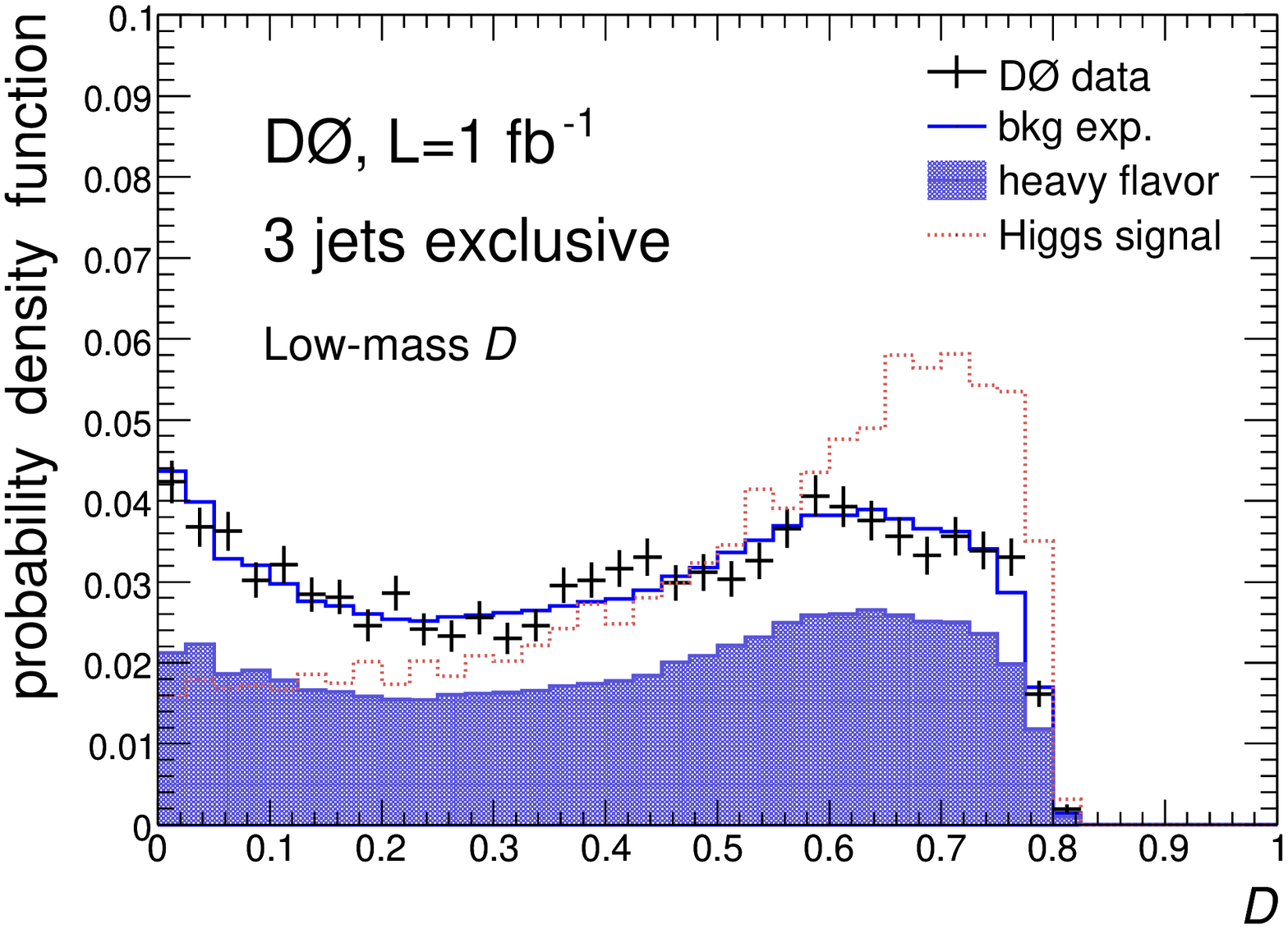}{Comparison of the low-mass likelihood distributions for the 3Tag data and background (bkg exp.) defined by Eq.~\ref{eq:BkgModel}. Every event has two entries. Black crosses refer to data, the solid line shows the total background estimate, and the shaded region represents the heavy flavor component ($b\bar{b}b$, $b\bar{b}c$, and $c\bar{c}b$). The distribution for a Higgs boson of mass 100~\gevcc is also shown.}{fig:ThreeTagModel_YLH_MH100}

\begin{figure*}
   \bc

   \includegraphics[width=5.5cm]{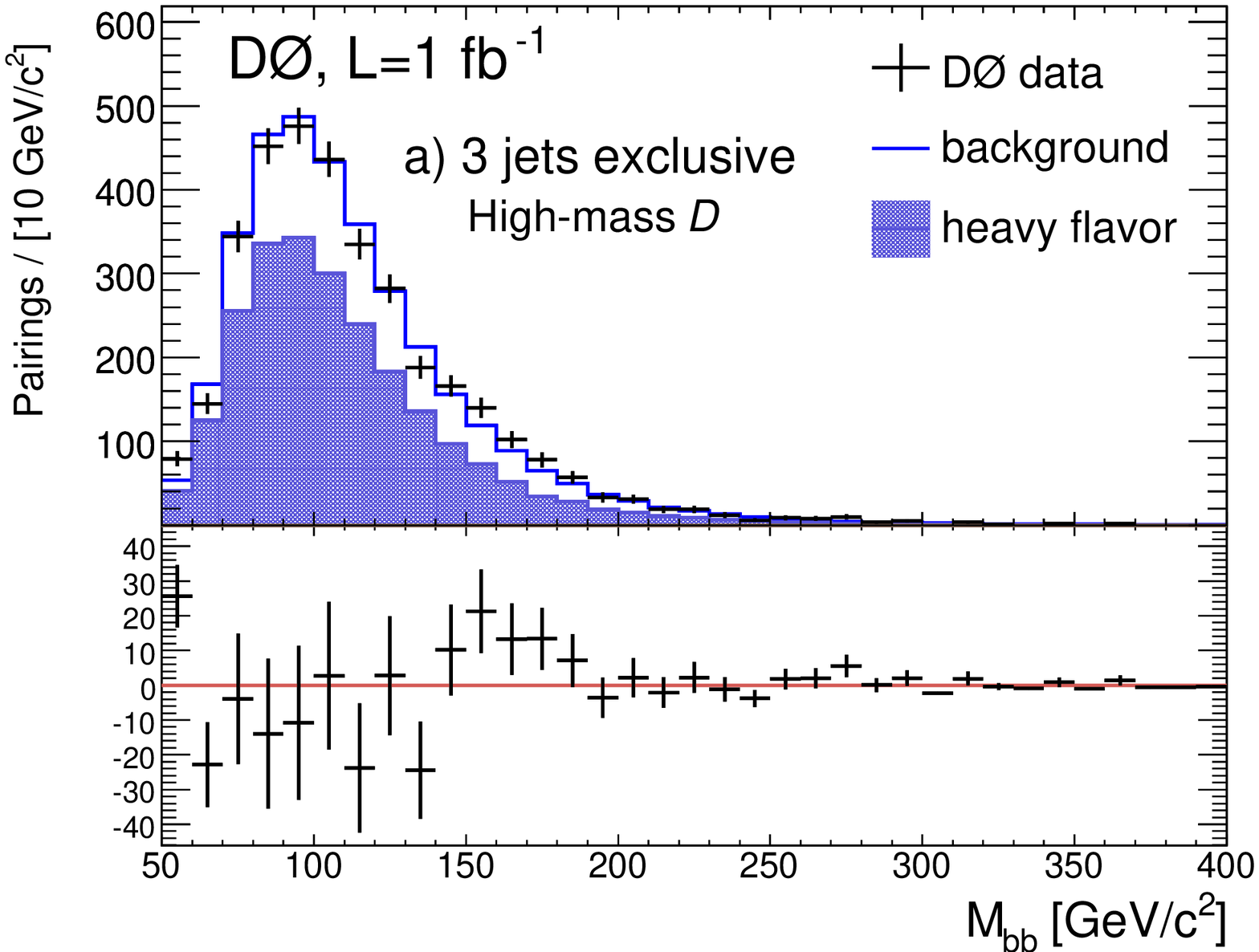}\vspace{1.mm}
   \includegraphics[width=5.5cm]{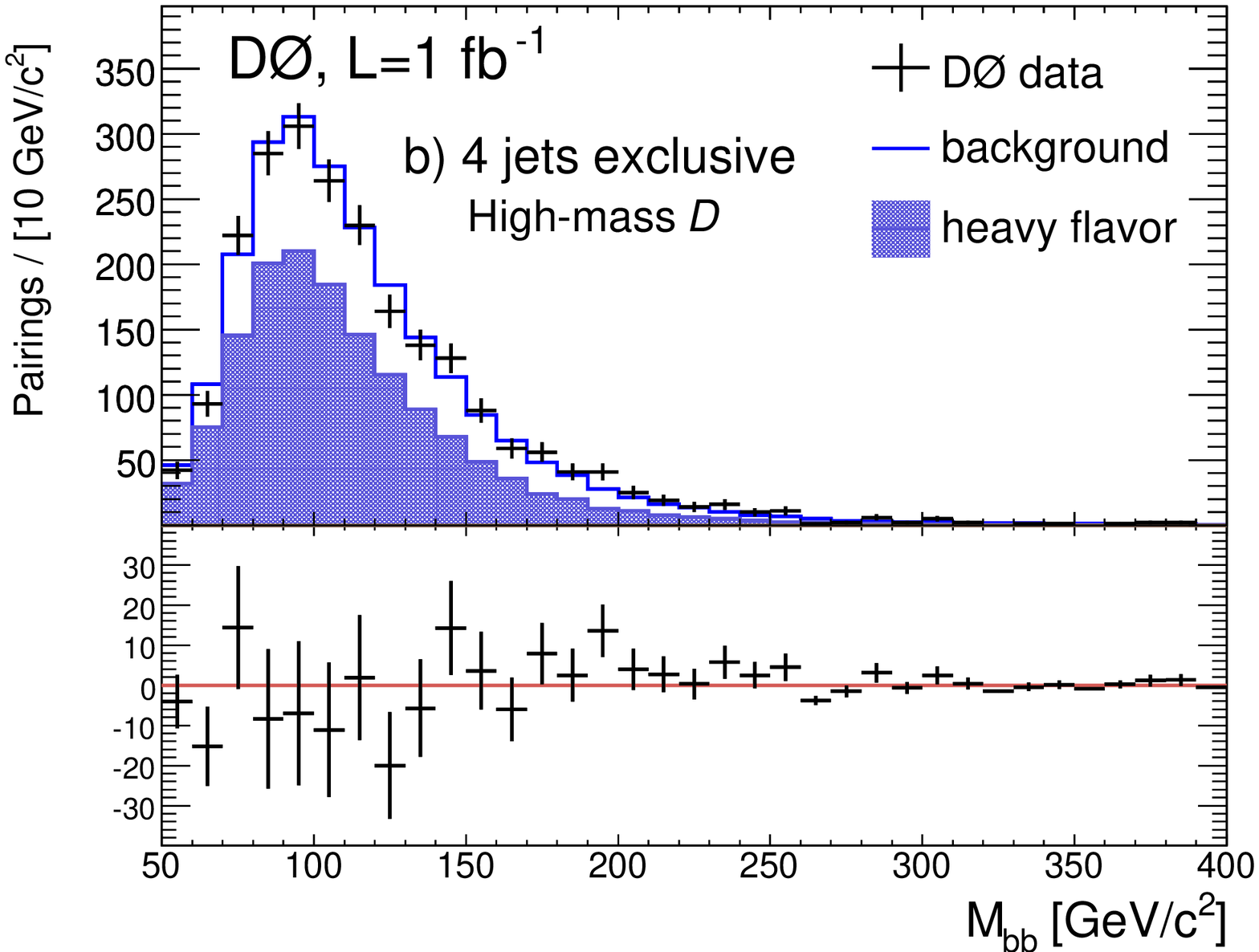}\vspace{1.mm}
   \includegraphics[width=5.5cm]{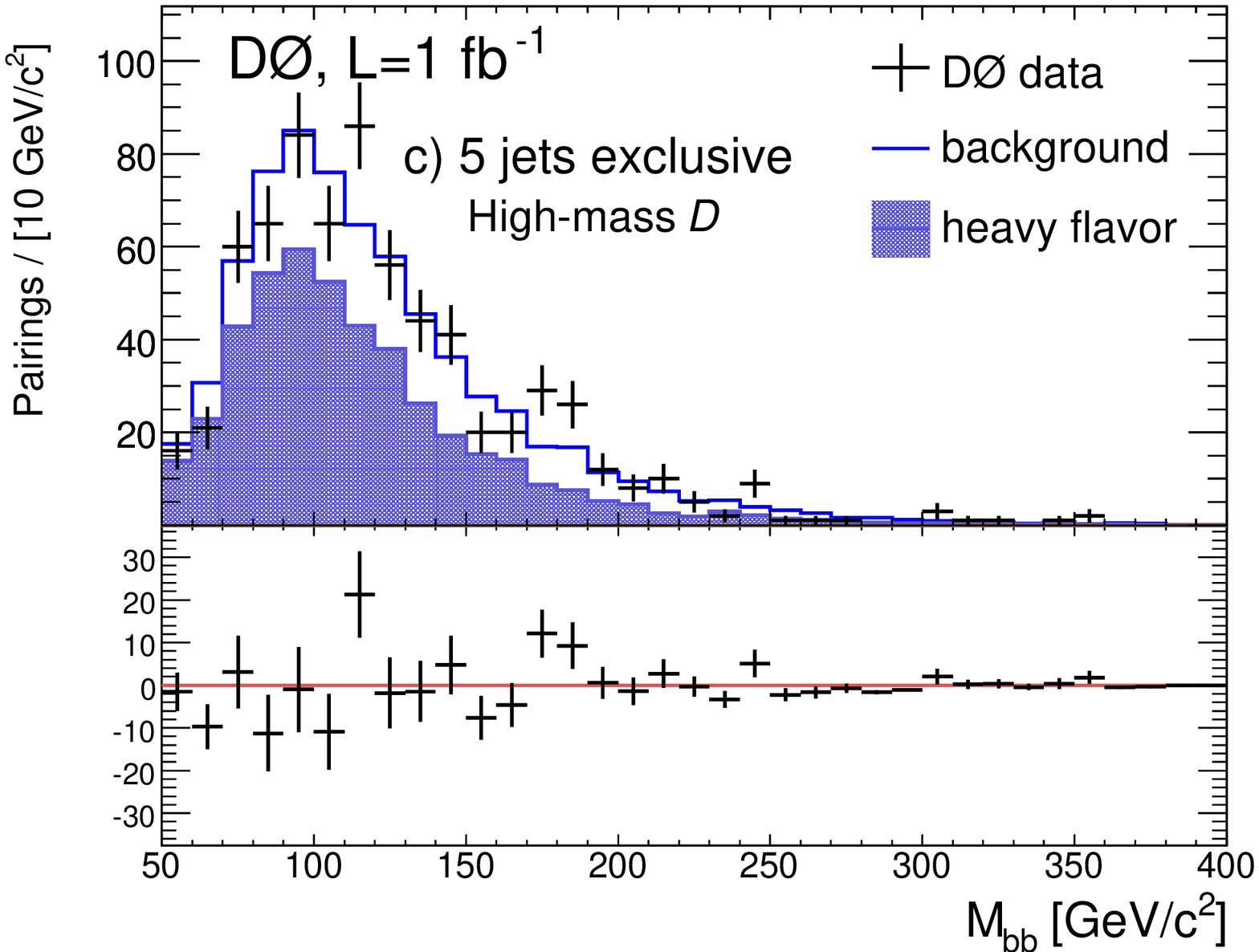}
   \caption{Invariant mass for the high-mass likelihood region for the exclusive a) three-jet b) four-jet, and c) five-jet channels. Black crosses refer to data, the solid line shows the total background estimate, and the shaded region represents the heavy flavor component ($b\bar{b}b$, $b\bar{b}c$, and $c\bar{c}b$). The lower panels show the difference between the data and the background expectation.}
   \label{fig:HighLikelihoodControlRegion_AbsMass}
   \ec
\end{figure*}

The selection cuts on $\mathcal{D}$, $b$-tagging, and number of jet-pair combinations per event are optimized by maximizing the expected sensitivity. The optimal cuts for the likelihood vary between 0.25 and 0.60 depending on the jet multiplicity and Higgs boson mass. The agreement of the data and the background expectation is verified in a control region where the impact of any Higgs boson signal is limited, defined by $\mathcal{D} < 0.25$. The agreement is also verified in the case when no likelihood cut is applied.
Figure~\ref{fig:HighLikelihoodControlRegion_AbsMass} shows the invariant mass for the optimized high-mass likelihood cuts.

Several sources of systematic uncertainties affecting the background shape through the ratio $\frac{S_{\rm 3Tag}^{MC}(\mathcal{D},M_{bb})}{S_{\rm 2Tag}^{MC}(\mathcal{D},M_{bb})}$ in Eq.~\ref{eq:BkgModel} are considered. The dominant uncertainty, due to the background composition, is estimated by varying the ratio of $b\bar{b}j$ and $b\bar{b}b$ events in the sample corresponding to the uncertainties from the background composition fit. In addition, the smaller uncertainties from the kinematic dependence of the $b$-tagging, the $b$ jet energy resolution, the $b\bar{b}b$ and $b\bar{b}j$ kinematics, and finally the trigger-level $b$-tag requirement were also evaluated and included in the systematic uncertainty on the background shape.

The Modified Frequentist method~\cite{cls} is used to estimate the compatibility of the data with the background-only hypothesis ($1-CL_b$) as well as to derive limits at the $95$\% C.L. on the cross section times branching ratio as a function of $m_A$. Only the shapes (not the normalization) of the $M_{bb}$ distributions are used to discriminate signal from background, assuming the width of $\phi$ to be narrow relative to the experimental resolution. Table~\ref{tab:results} shows the limits and the $1-CL_b$ values obtained versus the hypothesized Higgs boson mass.
\begin{table}[!h]
  \bc
 \begin{tabular}{l r r r }   \hline \hline  
 Mass                 &\hspace{0.25cm} $\sigma\times \BR$ &\hspace{0.25cm} $\sigma\times \BR$  &\hspace{0.25cm} $1-CL_{b}$   \\   
 ($\mathrm{GeV/c}^2$) &\hspace{0.25cm} Exp.(pb)           &\hspace{0.25cm} Obs.(pb)            &\hspace{0.25cm} (in \%)\\  \hline  

    90  & $ 170 ^{+ 72}_{ - 52}$ & 184&       39  \\
    100 & $ 117 ^{+ 48}_{ - 35}$ & 128&       38 \\
    110 & $  71 ^{+ 29}_{ - 20}$ &  69&       52 \\
    120 & $  41 ^{+ 18}_{ -  9}$ &  34&       73 \\
    130 & $  28 ^{+ 12}_{ -  7}$ &  24&       70 \\
    140 & $  25 ^{+ 11}_{ -  6}$ &  22&       60 \\
    160 & $  17 ^{+  8}_{ -  4}$ &  26&       12 \\
    180 & $  13 ^{+  5}_{ -  4}$ &  23&      4.4 \\
    200 & $   9 ^{+  4}_{ -  3}$ &  17&      7.0 \\
    220 & $   7 ^{+  3}_{ -  2}$ &  12&       12 \\

\hline \hline \end{tabular}

   \caption{Cross section limits as a function of Higgs boson mass. Columns two and three show the expected (Exp.) and observed (Obs.) limits on the cross section times branching fraction to $b\bar b$. The total one-sigma uncertainty on the expected limits is also displayed. The last column shows the value of $1-CL_b$.}
  \label{tab:results}
  \ec
\end{table}
The low $1-CL_b$ values around a Higgs mass of $180$~\gevcc\ are due to a slight excess over the expected SM background.

The results of this search can be used to set limits on the parameters of the MSSM. As a consequence of the enhanced couplings to $b$ quarks at large $\tan\beta$ the total width of the neutral Higgs bosons also increases with $\tan\beta$. This can have an impact on our search if the width is comparable to or larger than the experimental resolution of the reconstructed invariant mass of a di-jet system. To take this effect into account, the width of the Higgs boson is calculated with {\sc feynhiggs}~\cite{feynhiggs} and included in the simulation as a function of the mass and $\tan\beta$ by convoluting a relativistic Breit-Wigner function with the NLO cross section. In the MSSM the masses and couplings of the Higgs bosons depend, in addition to $\tan\beta$ and $m_A$, on the SUSY parameters through radiative corrections.  Limits on $\tan\beta$ as a function of $m_A$ are derived for two particular scenarios assuming a CP-conserving Higgs sector~\cite{Carena:2005ek}: the $m^{\rm max}_{h}$ scenario and the no-mixing scenario. Since the results depend considerably upon the Higgs sector bilinear coupling $\mu$, its two possible signs are also probed.

\begin{figure}[!h]
  \bc 

 \includegraphics[width=2.8cm]{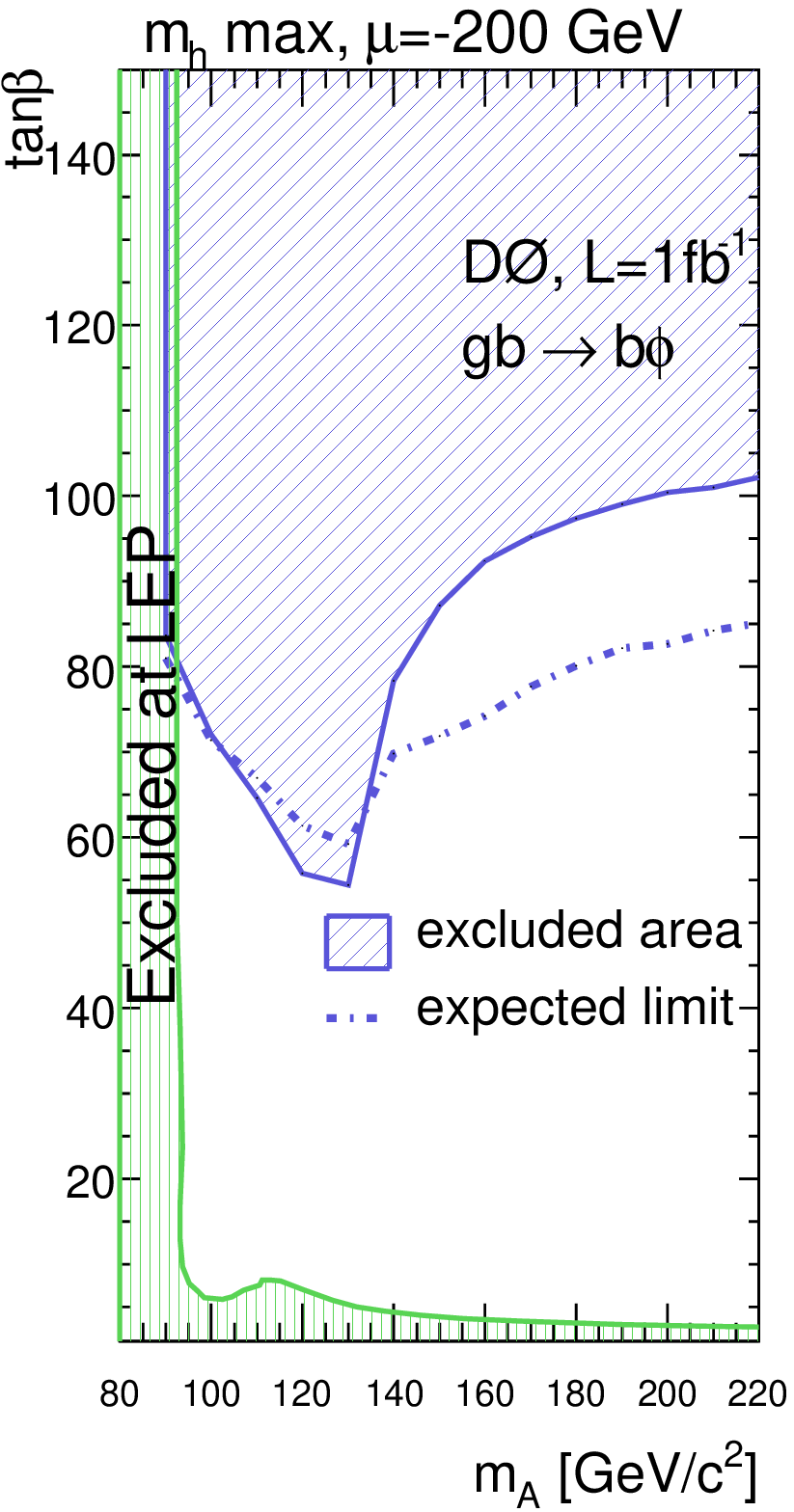}
 \includegraphics[width=2.8cm]{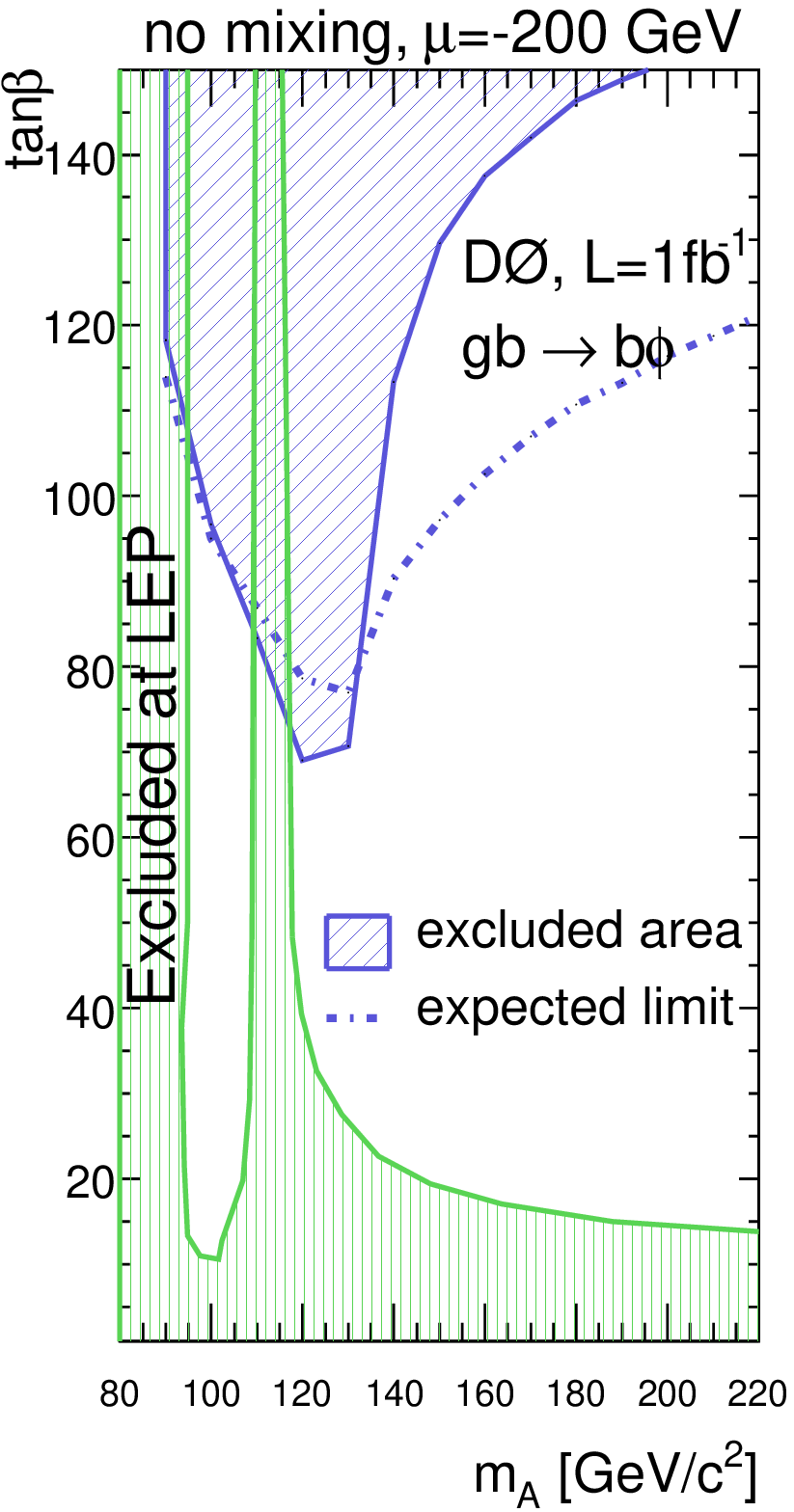}
 \includegraphics[width=2.8cm]{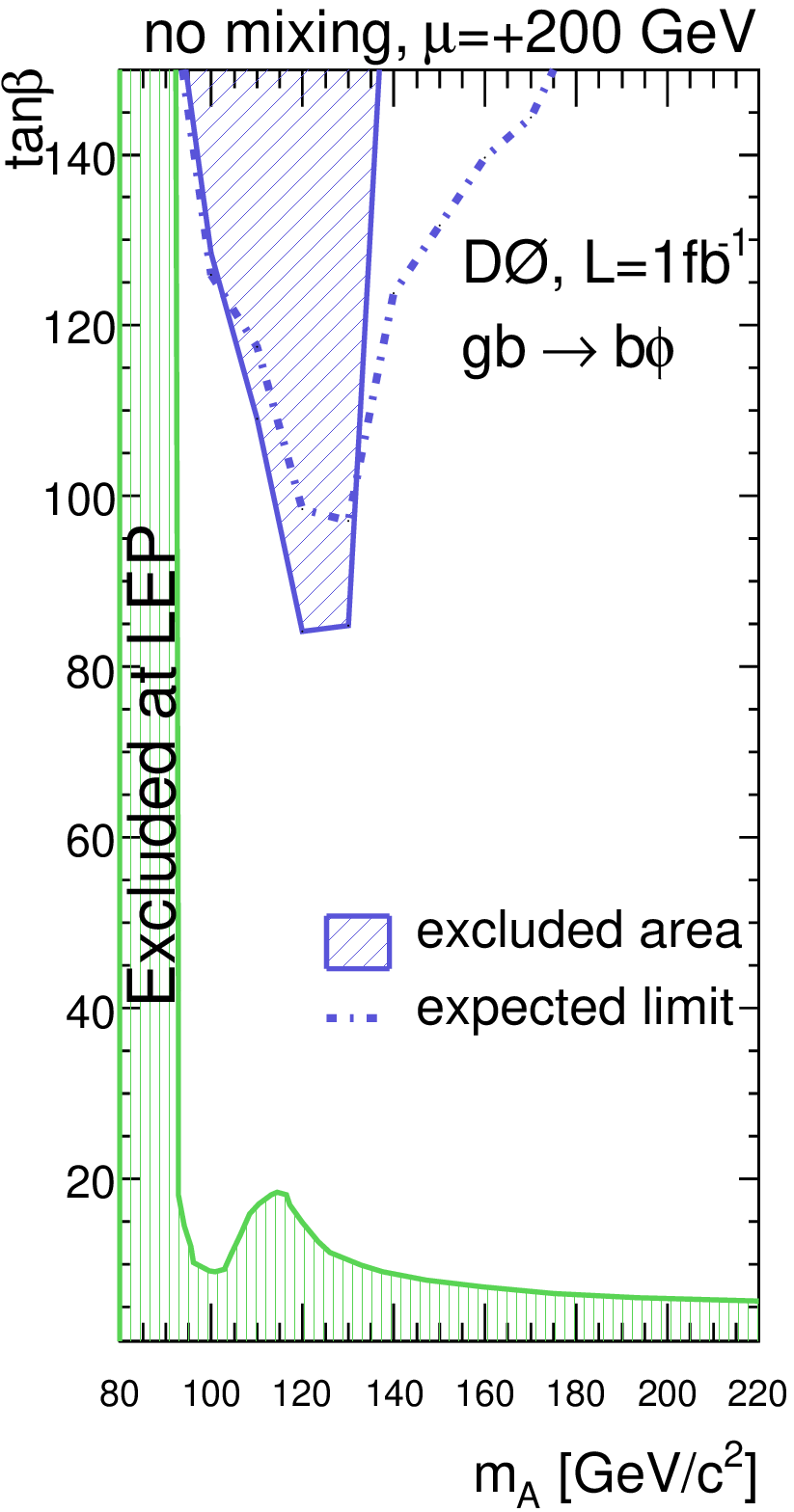}

   \caption{95\% C.L. exclusion limits in the $(m_A,\tan\beta)$ plane for $m^{\rm max}_{h}$, $\mu= -200$ GeV, and no-mixing, $\mu=-200$ GeV and $\mu=+200$ GeV. The exclusions from LEP are also shown \cite{cite:LEP_exclu}. The width of $\phi$ is larger than $70$\% of $m_A$ above $\tan\beta = 100$ in the $m^{\rm max}_{h}$, $\mu= -200$ GeV scenario.}
  \label{fig:FinalLimit_nomix_mu_positive} \ec
\end{figure}

Figure~\ref{fig:FinalLimit_nomix_mu_positive} shows the results obtained in the present analysis interpreted in these different MSSM scenarios. Substantial areas in the MSSM parameter phase space up to masses of $200$\gevcc are excluded. No exclusion can be obtained for the $m^{\rm max}_{h}$, $\mu > 0$ scenario.  This analysis extends the mass range over which the search is performed. In addition these results benefit from NN $b$-tagging and a likelihood discriminant as well as improved modeling and a robust limit setting procedure, using only the shape of the discriminating variable.

%
We thank the staffs at Fermilab and collaborating institutions, 
and acknowledge support from the 
DOE and NSF (USA);
CEA and CNRS/IN2P3 (France);
FASI, Rosatom and RFBR (Russia);
CNPq, FAPERJ, FAPESP and FUNDUNESP (Brazil);
DAE and DST (India);
Colciencias (Colombia);
CONACyT (Mexico);
KRF and KOSEF (Korea);
CONICET and UBACyT (Argentina);
FOM (The Netherlands);
STFC (United Kingdom);
MSMT and GACR (Czech Republic);
CRC Program, CFI, NSERC and WestGrid Project (Canada);
BMBF and DFG (Germany);
SFI (Ireland);
The Swedish Research Council (Sweden);
CAS and CNSF (China);
and the
Alexander von Humboldt Foundation.
%

\end{document}